# Influence of the Cathodes Microstructure on the Stability of Inverted Planar Perovskite Solar Cells

Svetlana Sirotinskaya*, Roland Schmechel, Niels Benson

University of Duisburg-Essen, Faculty of Engineering, Institute of Technology for Nanostructures (NST), Building BA, Bismarckstr. 81, 47057 Duisburg, Germany

E-mail: niels.benson@uni-due.de

One of the main challenges for perovskite solar cells (PSC) is their stability, due to environment-induced perovskite decomposition. The resulting decomposition compounds are mobile and may, therefore, react with charge carrier extraction layers or the contact metallization, in addition to enhancing the recombination rate in the absorber layer. In this contribution, the influence of different contact metallization layers, such as aluminum (Al), silver (Ag), gold (Au) and nickel (Ni) on the storage stability of inverted planar methylammonium lead iodide (MAPI)-based perovskite solar cells without encapsulation has been investigated. For this study current-voltage (*J-V)* and impedance measurements in combination with scanning electron microscope (SEM) and Energy-dispersive X-ray spectroscopy (EDX) analysis were used to examine and correlate structural device information with the development of PSC electrical properties. While a strong perovskite decomposition and further iodide diffusion to the contacts were detected for devices using Al, Ag or Au as cathode electrodes, the microstructure of Ni cathodes inhibits such decomposition process. This experiment has allowed for the realization of MAPI based solar cells with Ni contacts, which exhibit no efficiency decrease below as-fabricated values for up to one month of storage and select AM1.5 testing in ambient atmosphere.

**Introduction**

Organic-inorganic perovskite materials such as methylammonium lead iodide (MAPI) have attracted great attention during the last years because of their high performance for photovoltaic applications. Their easy production from the solution, crystallization at low temperatures and rapid development of efficiencies from 3%[1] to over 22%[2] make this material class one of the most promising modern material systems for solar cells. This high performance efficiency is based on physical properties of perovskites, such as tunable band gap[3], good absorption properties in the solar spectral range[4] and formation of only energetically flat defect states in the band gap[5], which leads to a long diffusion length[6] and a large charge carrier lifetime[7]. However, the perovskite-based photovoltaic devices show mostly low stability due to moisture, high temperatures and light. The often-reported low stability of PSCs is caused by different degradation processes of the absorber material, as well as the corrosion of the extraction and contact layers. Under illumination and the exposure to moisture, MAPI decomposes in different iodine-containing components (e.g. MAI, $PbI_2$, $CH_3NH_3I$, HI, $I_2$ and MAI-$nI_2$)[8–11]. Such degradation not only impairs the quality of the absorber layer, but also leads to the reaction of the decomposition products with the extraction layers[12]. The consequences are additional energetic barriers at the extraction layer/perovskite and electrodes/extraction layer interfaces and thus the inhibition of the electronic transport. Moreover, such iodine components can react with the metal contacts due to the diffusion processes through electron (ETL) and hole (HTL) transport layers, which causes the observed corrosion of metals. To prevent the PSCs degradation, different approaches have been suggested. Several research groups use additional layers which inhibit the diffusion of compounds or encapsulate the solar cell devices to decelerate a decomposition of the absorber layer[13,14,15]. Other groups achieve highly stable solar cells[12,16,17] by their choice in electron transport layers. This method has shown its efficiency, but it is limited to an available selection of extraction layers and limits free choice, regarding the largest open-circuit voltage and hence the highest cell efficiencies. Due to this, several research groups have started to consider the contact metallization in order to increase the device stability without changing the PSC architecture. The modification of the contacting metals has the advantage, that it does not influence the open-circuit voltage[18]. It was shown, that Ag electrodes degrade due to the chemical reaction between decomposed perovskite iodide components and Ag electrodes[19,20,21,22]. A similar but stronger reaction is observed for Al contacts[10,23,24]. Another degradation process is observed when using Au contacts[25,26]. In this case, there is not only an iodine diffusion to the contact interface, but also an Au atom diffusion into the perovskite layer. This leads to an enhanced charge carrier recombination, which impairs the performance of the solar cells. Another issue with the use of Au contacts is a catalytic effect of Au with methylammonium lead iodide, as reported by Kerner et al.[27]. This contribution suggests the Au accelerated decomposition of MAPI to hydroiodic acid (HI). Recently Shlenskaya et al. have suggested even another degradation process for Au electrodes in contact with MAPI, based on the formation of reactive iodide melts and their interaction with Au[8,9]. Here, we investigate the PSC stability depending on the contact metallization. For our investigation, we chose Al, Ag and Au as standard contact materials for PSCs and Ni as a low cost, stable alternative to Au. We show, that the microstructure of evaporated metals has a significant influence on the decomposition processes in the perovskite layer and the usage the metals with high melting temperature can lead to the natural encapsulation of PSCs.

**Results**

The architecture of the investigated *p-i-n* planar PSCs and their energy level diagram is shown in Fig. 1 a), b). The following metal work function values were assumed: 4.6 eV (Ag), 4.2 eV (Al), 5.2 eV (Ni), 5.3 eV (Au)[28]. As HTL and ETL, copper iodide (CuI) and buckminsterfullerene (C60) were used, respectively. Bathocuproine (BCP) layer was used as an additional hole blocking layer. In order to investigate the influence of the different metallizations on the cell performance and their stability, the *J-V* measurements were carried out using an AM1.5 spectrum with a power density of 100 mW/cm$^2$. The characterization was done directly after the preparation (figure 1 c) – f)) and during the next 32 days (figure 3). For the investigation of the device stability, the samples were stored in the dark at 20°C-23°C in air (about 20%-30% RH) and under inert nitrogen ($N_2$) atmosphere (<0.1 ppm $O_2$, $H_2O$). Figure 1 d) – f) illustrates the dependency of the average solar cell power conversion efficiencies (*PCE*), open-circuit voltages ($V_{oc}$) and short-circuit current densities ($J_{sc}$) on the metal work function. The best-achieved efficiencies are about 16% for devices

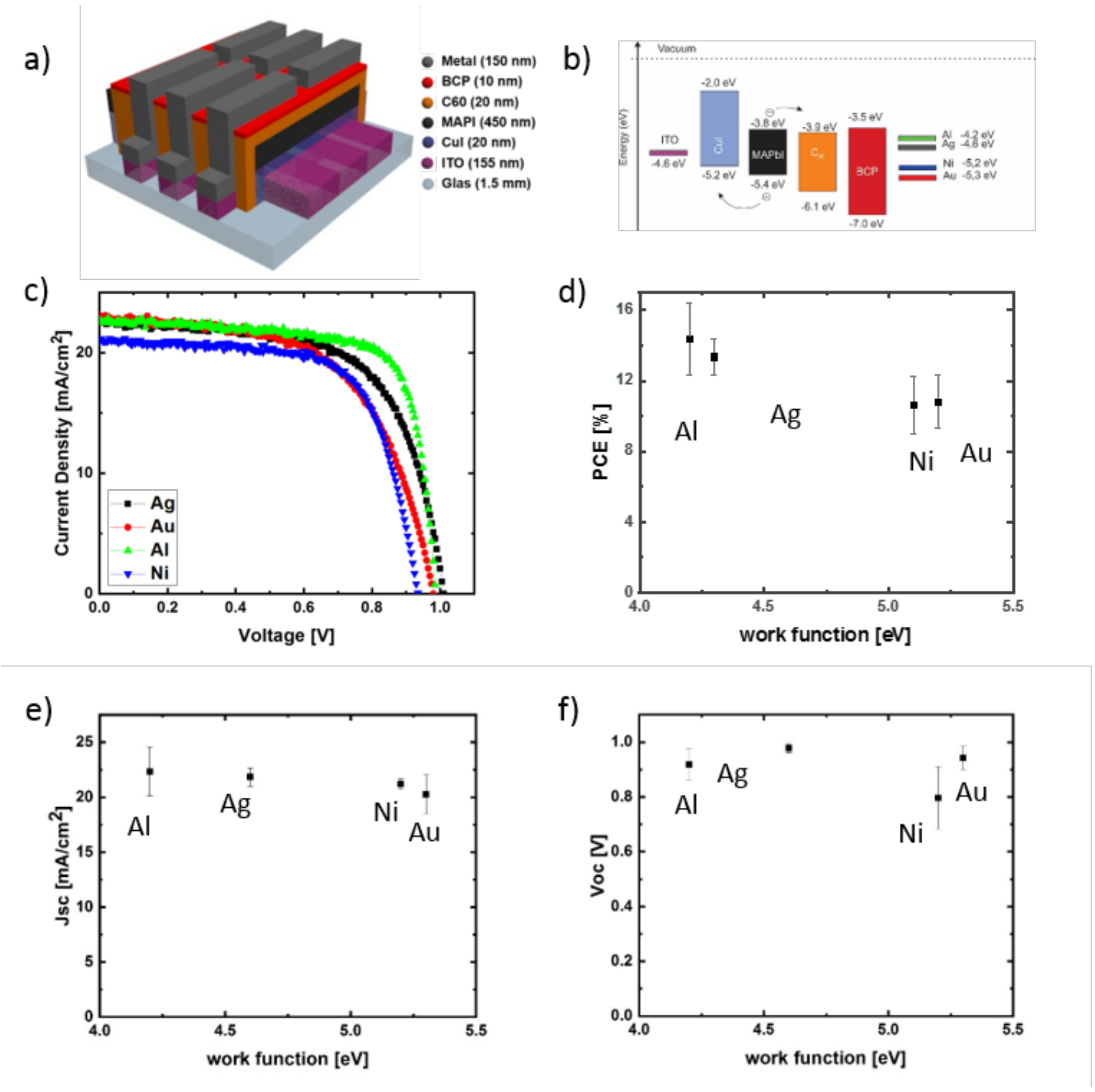

**Figure 1. a) Device architecture of a planar PSCs, b) band diagram of layers in a PSC, c) *J-V* characteristics of typical fresh devices with different contact materials, d) dependency of average efficiency, e) open-circuit voltage and f) short-circuit current on the work function of contact metals. For each metallization 36 PSCs were tested.**

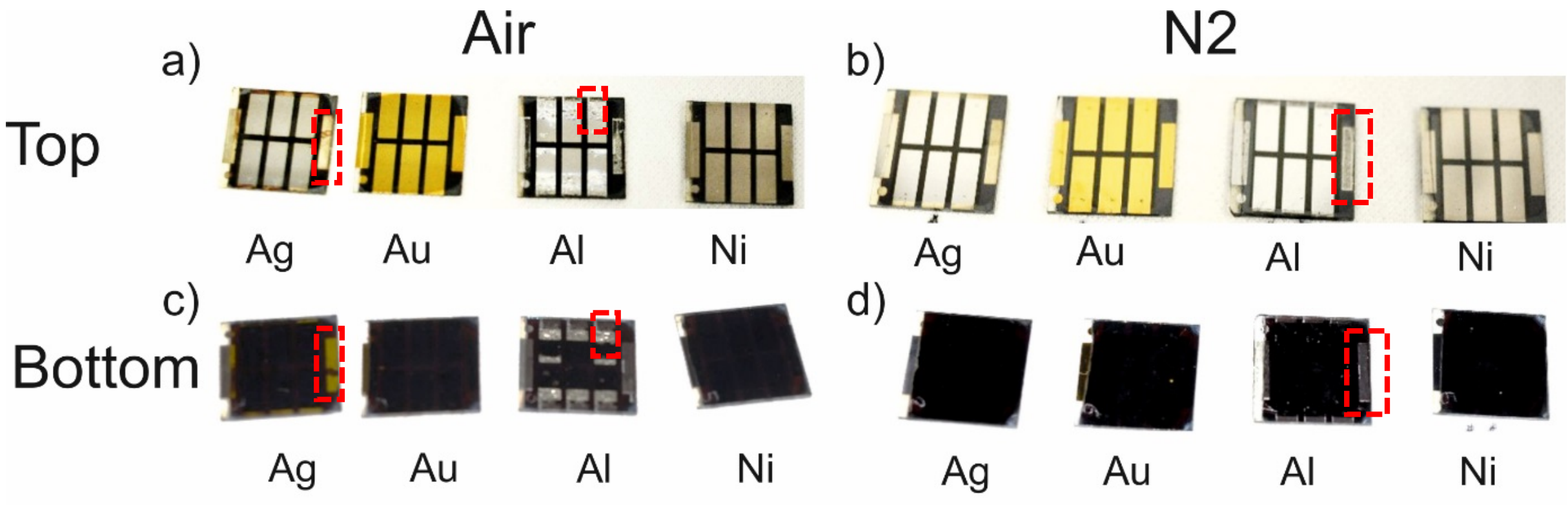

**Figure 2. a), b) Top, and c), d) bottom view of solar cells after one-week storage a), c) in the ambient and b), d) in $N_2$ atmosphere with Ag, Au, Al and Ni electrodes**

with Al cathodes and are comparable with the highest efficiencies of the devices with the same charge-carrier transport layers and device architecture[17,29]. Since the aim of this contribution was the influence investigation of the top metallizations on the PSC stability no further charge transport layers were tested. As shown in figure 1 f) no strong correlation between $V_{OC}$ and the metal work function was observed. This matches well with the results of Behrouznejad et al. [18], which were obtained using a regular *n-i-p* structure. For Al, Au and Ag a $V_{oc}$ between 0.9 V and 1 V was determined, while the $V_{oc}$ for Ni with a value of 0.8 V was clearly reduced. We suggest this to be a consequence of the energetic barrier, caused by a non-ohmic contact between the metal and the BCP layer[30,31]. In figure 2 top- and bottom-view images of the solar cells with the different top contact metallizations, after one week of testing and storage in air and $N_2$ are illustrated. We note that all the devices were stored in dark, so that the light-induced degradation was minimized. The first indication for the degradation is directly visible for cells with Al and Ag contacts. The solar cells with Al contacts show strong degradation after storage in air (figure 2 c)) and visible degradation in $N_2$ (figure 2 d)). The metal contacts become visible through the black perovskite layer. This indicates a strong Al corrosion process, making further measurement impossible. The Ag contact degradation and MAPI decomposition for cells stored in air is noticeable by the yellow color at the contact edge (top view) and of the perovskite below this contact (bottom view).

Dashed red lines in figure 2 mark all visible degradation effects. No visible degradation of PSCs with Au and Ni electrodes is observed.

The result of the electrical degradation evaluation in the respective atmospheric conditions over the entire time period of 32 days is illustrated in figure 3, for the parameters *PCE*, $V_{OC}$, $J_{SC}$ and *FF*. These values have been normalized to their initial values obtained for the as-fabricated devices ($t$=0). While the solar cells with aluminum and silver contacts degrade much more rapid in air than in nitrogen, the degradation of gold electrode cells is almost independent of the storage conditions. This result matches well with the optical observations discussed for figure 2. Remarkably, strong improvement of approximately 50% in *PCE* was obtained during the first days of testing for the solar cells with Ni top electrodes. More importantly, even after the controlled aging for 32 days in $N_2$ and ambient atmosphere, the *PCE* does not drop below the *PCE* values obtained at $t$=0. This indicates a very slow aging of these devices. As already mentioned, the PSCs with Al contacts underlie the most rapid degradation, which is mainly a consequence of a drop in $J_{SC}$ during the first days. For this type of cell, an efficiency decrease of 100% and 90% was obtained after two-week storage in ambient and nitrogen atmosphere. This is in line with the observed strong electrode corrosion and perovskite decomposition. The cells with Ag metallization show a clear difference in their degradation behavior depending on the storage conditions. While the cells stored for two weeks in $N_2$ degrade only by approximately 10% in *PCE* due mainly to a $V_{OC}$ decrease, cells stored in ambient show a 60% degradation in *PCE* due to a decrease in $V_{OC}$, $J_{SC}$ and *FF*. A degradation in all *J-V* parameters was observed for PSCs with Au electrodes. We observed an efficiency decrease of ~50% after one-month storage, independent of the storage conditions. For the special case of the cells with Ni top electrodes, the gain in *PCE* during the first days of testing is the result of an increase in open-circuit voltage. This is in line with the observation by Jiang et al., who observed a $V_{OC}$ and *FF* enhancement in MAPI *n-i-p* devices with Ni electrodes after one day of storage in a desiccator[32]. We suggest this could be the result of a slow diffusion of Ni atoms into the BCP layer and generation of transport states in the BCP band gap[30,31]. Such a process may lead to a Ni/BCP interface improvement. We also speculate that the other reason for the low initial $V_{OC}$ and its further enhancement could be a consequence of the slightly damaging of BCP during the Ni evaporation process and its further self-healing. As already implied above, no significant degradation in solar cell performance for MAPI devices with Ni top electrodes was observed during the 32 days of degradation testing.

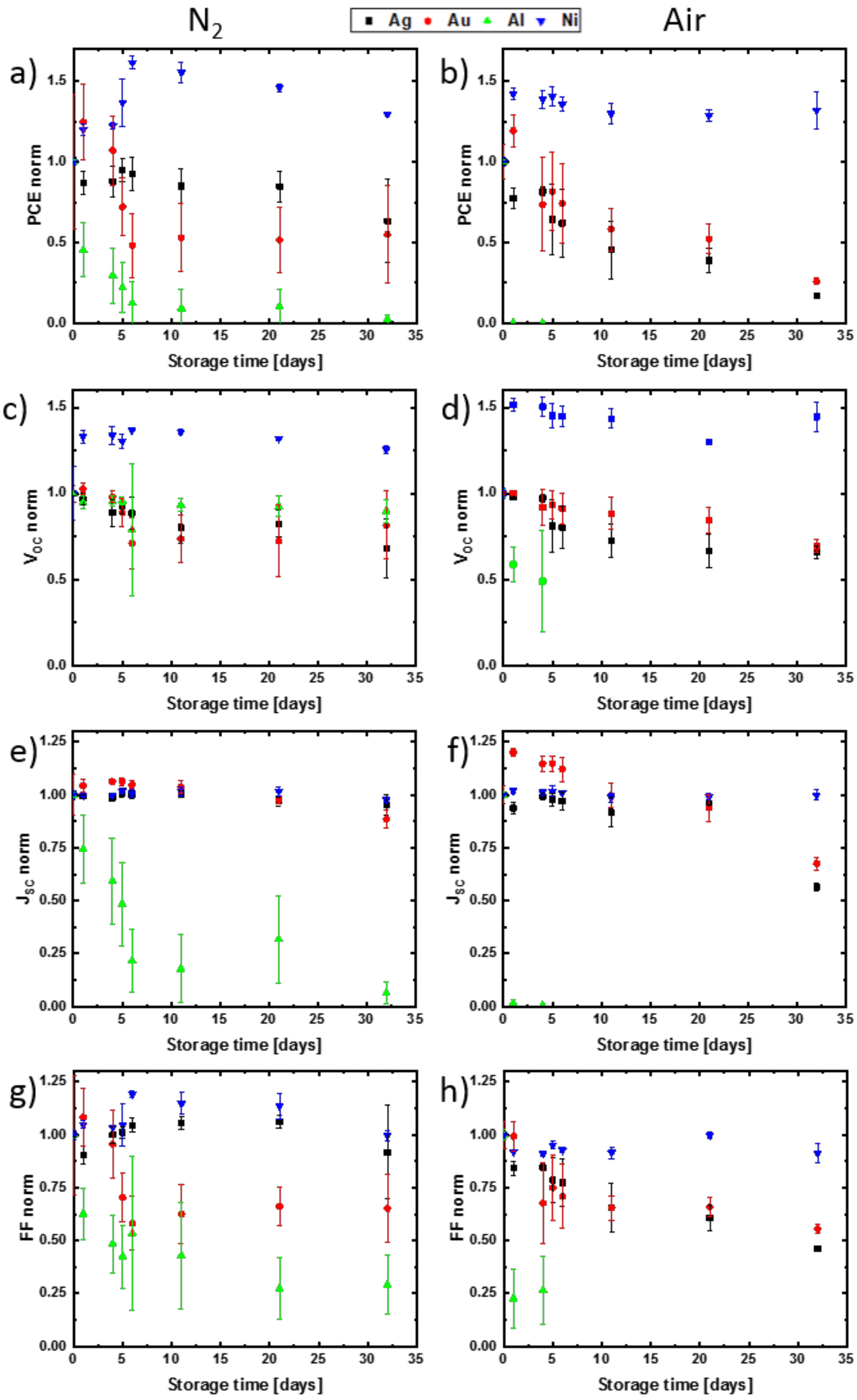

**Figure 3. Degradation of PSCs due to the storage in ambient (a-d) and nitrogen (e-h) atmosphere. Dependence of normalized *PCE* a), e), $V_{OC}$ b), f), $J_{SC}$ c), g) and *FF* d), h) on the storage time.**

In order to further investigate the observed contact corrosion, impedance spectroscopy measurements (IS) were carried out. This method allows to distinguish between the influence of the perovskite bulk and the ETL/perovskite interface on the electronic PSC properties. This is the consequence of a different frequency response for different charge transport mechanisms[33]. While the response of ion and deteriorated electron transport in degraded devices is measurable at low frequencies, the influence of the perovskite/ETL interface on the charge carrier transport corresponds to the high-frequency part[3333]. For this reason, the frequency range from 1kHz to 1MHz was chosen. The measurements were done under 100 mW/cm$^2$, AM 1.5 illumination and under open-circuit condition, ensuring, that the device operates under maximum recombination conditions[34,35]. Moreover, under such condition, the charge carriers are accumulated at the perovskite/ETL interface, increasing the recombination probability at these interfaces[34]. Here, the two-component Voight model is used for the data interpretation, as outlined in the supplementary information (SI)[36]. In this model, the interaction between the electrode or ETL with the perovskite layer is represented by the semi-circle obtained between 1MHz and 10kHz, which is shown in figure 4 for the PSCs with different top contact metallization[18,37]. Because of the assumed diffusion of metals into the ETL, its small thickness and the fact that the same ETL was used for all PSCs, we suggest the observed differences in IS measurements to be caused by the different contact metallizations. While the observed change in pure metal series resistance is almost neglectable, which would be represented by a semicircle shift on the Z´-axis, a significant increase in the semi-circle radius was observed. This implies a significant increase in interfacial charge-carrier recombination activity (due to interface defect state formation, etc.), which will be the cause for the discussed *J-V* characteristic change during the aging experiment. After two weeks of storage in air for the case of Ag electrodes, the increase in semi-circle radius corresponds to a perovskite/ETL interface resistance ($R_{per/ETL}$) increase by a factor of 20, while the perovskite/ETL interface capacitance ($C_{per/ETL}$) remains almost unchanged. During the same time frame, $R_{per/ETL}$ changed by a factor of 6 and 2 for PSCs with Au and Ni electrodes, implying a significantly slower metal/perovskite interface trap formation when compared to cells with Ag electrodes. This observation is supported by an SEM cross-sectional analysis of the different interfaces, demonstrating an Ag/MAPI interface deterioration, whereas no morphology change was detected for degraded PSCs with Au and Ni contacts (see SI). This analysis is outlined in the experimental section.

In contrast to the PSCs with Ag electrodes, almost no degradation of the contact interface is observed for impedance measurements of PSCs with Au and Ni electrodes. In order to investigate the degradation processes observed in the *J-V* characteristic of this cell type and to clarify the difference in the aging results for Au and Ni contacts, cross-section EDX measurements of the two-month-old devices stored in ambient atmosphere were carried out. While the mapping measurements (see SI) implicate a constant perovskite composition along the cross-section, line scan EDX measurements were carried out to further quantify the analysis (figure 5).

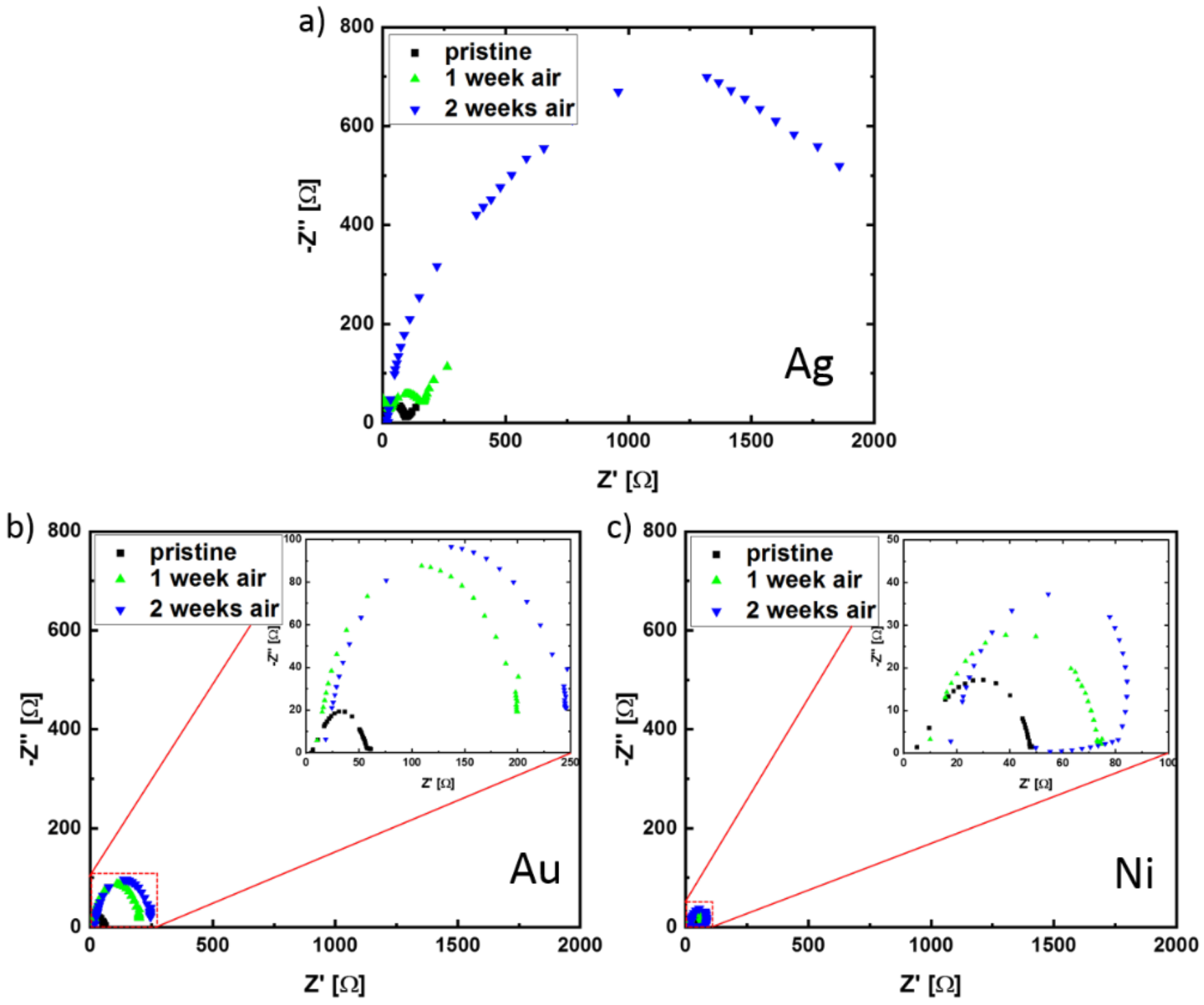

**Figure 4. Nyquist plots for solar cells with a) Ag, b) Au and c) Ni contacts directly after fabrication and after one- and two-week storage in air**

For the measurements, an extraction voltage of 15 kV and a probe current of 10 mA was used to detect the MAPI constituents iodide and lead. Because of the high extraction voltage and large detection depth, only approximate trends and average values can be considered for this type of investigation. The result of the experiment is illustrated in figure 5. It is surprising to see, that after two weeks of storage, the Pb:I ratio has degraded throughout the cell to a ratio of 1:1 for devices with Au and to a ratio of 1:1.2 for devices with Ag contacts. This stands in contrast to the result obtained for devices with Ni electrodes, where the Pb:I ratio remains 1:3 throughout the perovskite thin film, as it would be expected for an intact perovskite stoichiometry. We suggest the observed degradation in Pb:I ratio to be the consequence of a perovskite decomposition into volatile iodide compounds and their diffusion out of the perovskite layer/cell. This leads to the question, why is there such a pronounced difference between Ni contacted cells in comparison to the devices with Ag and Au electrodes. We consider this to be the result of the water/moisture concentration in the thin film as outlined by Huang et al.[11]. Depending on the metal contact morphology, moisture may diffuse into the perovskite layer not just from the sides of the cell, as suggested by Kato et al.[19], but also through the metal contact itself, which will result in an enhanced water uptake. The morphology of the evaporated Al, Ag, Au and Ni contact layers is shown in figure 6, using images obtained from an SEM analysis. It stands out, that Al, Ag and Au films have a clear polycrystalline structure with a broad crystallite size distribution, rough surface and well-defined grain boundaries, while the nickel film consists of highly dense nanocrystallites (<10 nm) and a smooth surface. The blurred structure under the nickel film is related to the perovskite surface under the metal layer. The origin of this morphological variation is the difference in melting temperature for the used metals[38,39,40,41,42]. It is well known, that thermally evaporated, as well as sputtered metal thin films have a polycrystalline morphology. Here the crystallite size is dependent on the ratio of the substrate temperature ($T_S$)/ metal melting temperature ($T_M$). Depending on this ratio, 4 different growth regions are possible as shown in figure 6. Zone 1 represents a nanocrystalline metal phase, while zone T yields a transition zone between the nanocrystalline structure of zone 1 and the larger grained polycrystalline zones 2 and 3. For the case at hand, the different metallizations were deposited without any cooling, at substrate temperatures of ~350 K. This results in a $T_S/T_M$ ratio of 0.36, 0.28, 0.26 and 0.2 for the respective Al, Ag, Au and Ni metallization. This means that Ni, Au and Ag will form thin films related to zone T. However, as zone T is the transition zone between zones 1 and 2, and by taking the SEM micrographs shown in figure 6 into account, we suggest, that the Ni thin films form more at the nanocrystalline end of the scale and exhibit undefined grain boundaries, while the Ag and Au thin films form a polycrystalline thin film with well-defined grain boundaries. As the grain boundary diffusion in metal films is usually several times larger than the lattice diffusion, the absence of clearly formed grain boundaries in the Ni thin films prevents moisture diffusion into the perovskite thin film[43,44]. This supports our hypothesis as outlined above, and results in a reduced MAPI decomposition rate, which translates into the observed higher stability of the photovoltaic device. Finally, we would like to point out, that the reason for the difference in degraded Pb:I ratio for the devices with Ag and Au metallization, is most likely the catalytic effect of gold on the decomposition of MAPI as discussed by Kerner et. al[27].

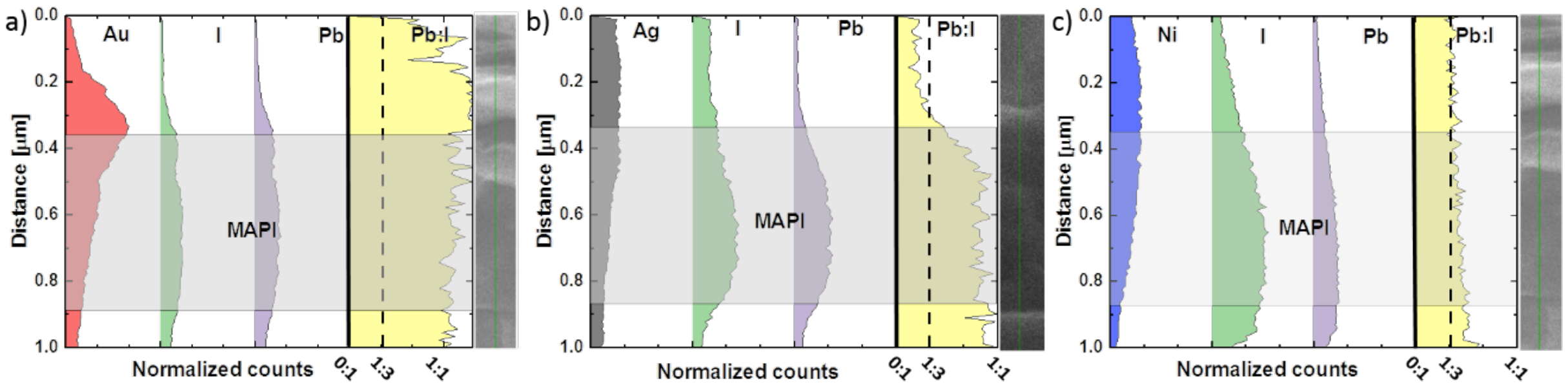

**Figure 5. Normillized cross-section EDX line scan of degraded PSCs with a) Au, b) Ag and c) Ni contacts. Iodide and lead are presented by green and purple color respectively.**

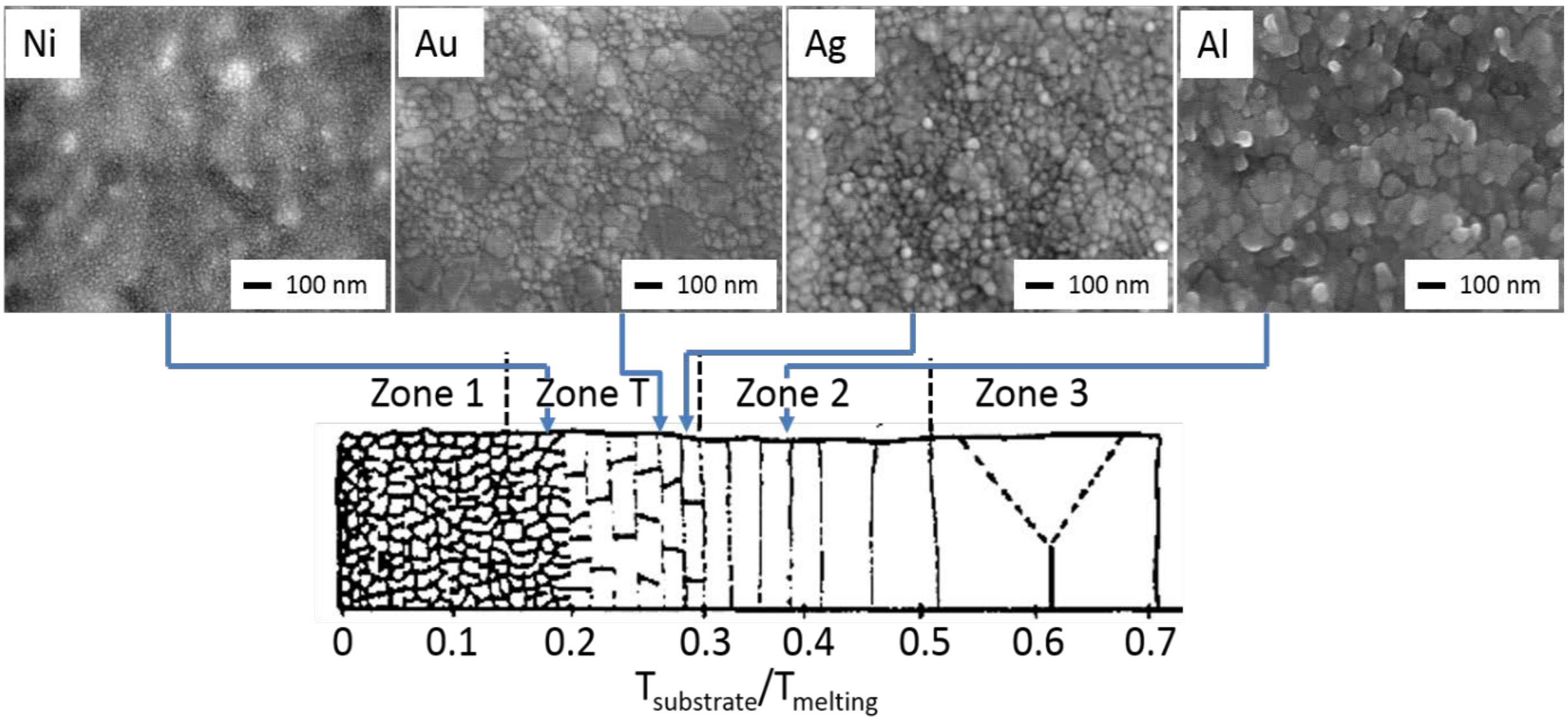

**Figure 6. SEM top view micrographs of 100 nm thick Ni, Au, Ag and Al layers. While the Ag and Au films have a clear polycrystilline structure whith a broad cristillite size distribution, rough surface and well defined grain boundaries, the nickel film consists of highly dense nanocrystillites (<10 nm) and has a smooth surface. This observation is in line with zone models, where the morphological differences are defined due to the substrate temperature ($T_S$)/ metal melting temperature ($T_M$) ratio. Modified from Ref. [40], with the permission of Elsevier.**

**Conclusion**

We have investigated the stability of inverted planar MAPI solar cells in dependence on the top contact metallization, using *J-V*, impedance as well as SEM and EDX analysis. We demonstrate that solar cell degradation is contact material dependent. While solar cells with Ag, Al and Au top electrodes degrade due to corrosion or decomposition effects, no cell degradation effects were observed for devices with Ni electrodes even after one month of storage and testing in $N_2$ or ambient atmosphere. This represents a strong exception from the rule. We demonstrate that this remarkable behavior is the consequence of two Ni contact properties. I. The low reactivity of Ni does not lead to contact interface corrosion as observed for solar cells with Ag or Al metallization. II. The highly dense packing of the Ni morphology, when compared to Ag, Al or Au, inhibits the water diffusion through the contact area, and consequently a strongly reduced water uptake. This effect strongly reduces the moisture-related MAPI decomposition and the formation of volatile Iodine containing groups for devices with Ni contacts.

**Acknowledgments**

The authors acknowledge financial support through the European Union in the Leitmarktwettbewerb NRW: Neue Werkstoffe, within the PeroBOOST project (EFRE-0800120; NW-1-1-040).

**Experimental**

Materials

MAI (98%), DMSO (>99,9% for molecular biology), DMF (99.8% anhydrous), Diethyl ether (99.9%), BCP (99.99%) and C60 (99.5%) were purchased from Sigma Aldrich. $PbI_2$ (99.9985%) was purchased from Alfa Aesar. Acetonitrile was received from HiPerSolv CHROMANORM. CuI was synthesized using the well-known method by the mixing of water solution of copper (II) sulfate with potassium iodide. Afterward, the sediment was washed with water and dried in vacuum.

Fabrication of solar cells for *J-V* and impedance characterization

Indium tin oxide (ITO)-coated glass sheets with a resistivity of 15 Ω sq $^{-1}$ were patterned by Naranjo substrates. They were subsequently cleaned with Aceton, a 3 % Mucasol solution, a 1:1 mix of Isopropanol and Acetone and finally Ethanol. Directly before applying the hole transport layer, remaining organic residues were removed by an oxygen plasma treatment for 15 min at 130°C. CuI solution was prepared by solving 10 mg of CuI powder in 1 ml Acetonitrile. Diethyl ether layer (about 30 nm) was deposited on the top of ITO substrates by spin-coating in a glovebox at 3000 rpm for 30 s. The CuI layer was then annealed for 10 min. at 100°C[17]. Afterward, the substrates were transferred out from the glovebox and the perovskite layer was deposited. We used the anti-solving method published by Park et al. to get a hole-free perovskite layer[45,46]. Perovskite solution was prepared by dissolving 1 mmol of $PbI_2$(461 mg) and MAI (156 mg) in 530 mg of DMF, in which 1 mmol of dimethyl sulfoxide (DMSO) (78 mg) was added. The solution was filtered using a syringe filter with a pore size of 0.45 µm. The 20 µl of solution was applied on the top of the CuI layer and then spin-coated at 4000 rpm for 25 s. After 12 s spin-coating, 0.28 ml of diethylether was added. The adduct film was converted to the polycrystalline MAPI layer by the annealing at 60°C for 1 min and at 100°C for 10 min subsequently. For the electron transport and hole blocking layer were used C60 (20 nm) and BCP (10nm), sequentially deposited by thermal evaporation under high vacuum (3*10-9 bar) was used. Finally, 150 nm of metals (Ag, Au, Al or Ni) was also thermally evaporated (at 2-8*10-9 bar) to create the top electrode.

Characterization

Solar Simulator

Photocurrent-voltage measurements were performed using a Keithley model 238 source measure unit controlled by a custom made LabView program. The voltage was varied between -0.1 and 1.1 V with a voltage step of 10 mV. The scan rate was 0.05 V/s. The solar simulator (Wacom WXS-155s-10) equipped with a 1000 W Xenon lamp, was used as a light source (AM 1.5 G), where the light intensity was accurately calibrated employing a pyranometer (CMP 11, Kipp & Zonen). The active area of all devices were 5 $mm^2$ and was defined by a shadow mask. For each material 36 solar cells were characterized.

Impedance measurements

The impedance spectroscopic measurements were performed with a Keithley model 4200-SCS at 30mV RMS amplitude at the open-circuit condition under 100 $mW/cm^2$ AM 1.5G irradiation. The frequency was varied between 1kHz and 1MHz and the AC perturbation voltage was 30 mV. Bevor impedance measurements, *J-V* characterization was performed, to find the open-circuit voltage values. The measurements were evaluated by the custom made MatLab program.

Scanning electron microscopy (SEM)

The morphological characterization of layers was made with a scanning electron microscope (Joel JSM 7500 F). Secondary electrons were detected. For the top view images, an acceleration voltage of 5kV and an emission current of 10 µA were used. For the cross-section images, the substrates were scratched with a diamond cutter and subsequently broken. The images were made in gentle beam mode with an acceleration voltage of 3 kV and decelerate voltage of 2 kV. The emission current was 10 µA.

Energy-dispersive X-ray spectroscopy (EDX)

The element analysis was carried out using EDX measurements on PSC cross-sections. A scanning electron microscope (Joel JSM 7500 F) with an EDX detector (Bruker 5030) was used. To achieve a sufficient count rate, the acceleration voltage was set to 15 kV and emission current to 10 µA.

# Referencies

1 A. Kojima, K. Teshima, Y. Shirai and T. Miyasaka, *Journal of the American Chemical Society*, 2009, **131**, 6050–6051.

2 W. S. Yang, B.-W. Park, E. H. Jung, N. J. Jeon, Y. C. Kim, D. U. Lee, S. S. Shin, J. Seo, E. K. Kim, J. H. Noh and S. I. Seok, *Science (New York, N.Y.)*, 2017, **356**, 1376–1379.

3 S. A. Kulkarni, T. Baikie, P. P. Boix, N. Yantara, N. Mathews and S. Mhaisalkar, *J. Mater. Chem. A*, 2014, **2**, 9221–9225.

4 S. de Wolf, J. Holovsky, S.-J. Moon, P. Löper, B. Niesen, M. Ledinsky, F.-J. Haug, J.-H. Yum and C. Ballif, *The journal of physical chemistry letters*, 2014, **5**, 1035–1039.

5 W.-J. Yin, T. Shi and Y. Yan, *Appl. Phys. Lett.*, 2014, **104**, 63903.

6 Q. Dong, Y. Fang, Y. Shao, P. Mulligan, J. Qiu, L. Cao and J. Huang, *Science (New York, N.Y.)*, 2015, **347**, 967–970.

7 Y. Bi, E. M. Hutter, Y. Fang, Q. Dong, J. Huang and T. J. Savenije, *The journal of physical chemistry letters*, 2016, **7**, 923–928.

8 N. N. Shlenskaya, N. A. Belich, M. Grätzel, E. A. Goodilin and A. B. Tarasov, *J. Mater. Chem. A*, 2018, **6**, 1780–1786.

9 A. A. Petrov, N. A. Belich, A. Y. Grishko, N. M. Stepanov, S. G. Dorofeev, E. G. Maksimov, A. V. Shevelkov, S. M. Zakeeruddin, M. Graetzel, A. B. Tarasov and E. A. Goodilin, *Mater. Horiz.*, 2017, **4**, 625–632.

10 E. M. Sanehira, Tremolet de Villers, Bertrand J., P. Schulz, M. O. Reese, S. Ferrere, K. Zhu, L. Y. Lin, J. J. Berry and J. M. Luther, *ACS Energy Lett.*, 2016, **1**, 38–45.

11 J. Huang, S. Tan, P. d. Lund and H. Zhou, *Energy Environ. Sci.*, 2017, **10**, 2284–2311.

12 H.-S. Kim, J.-Y. Seo and N.-G. Park, *J. Phys. Chem. C*, 2016, **120**, 27840–27848.

13 H. Back, G. Kim, J. Kim, J. Kong, T. K. Kim, H. Kang, H. Kim, J. Lee, S. Lee and K. Lee, *Energy Environ. Sci.*, 2016, **9**, 1258–1263.

14 A. Guerrero, J. You, C. Aranda, Y. S. Kang, G. Garcia-Belmonte, H. Zhou, J. Bisquert and Y. Yang, *ACS nano*, 2016, **10**, 218–224.

15 F. Igbari, M. Li, Y. Hu, Z.-K. Wang and L.-S. Liao, *J. Mater. Chem. A*, 2016, **4**, 1326–1335.

16 Y. Wei, K. Yao, X. Wang, Y. Jiang, X. Liu, N. Zhou and F. Li, *Applied Surface Science*, 2018, **427**, 782–790.

17 W.-Y. Chen, L.-L. Deng, S.-M. Dai, X. Wang, C.-B. Tian, X.-X. Zhan, S.-Y. Xie, R.-B. Huang and L.-S. Zheng, *J. Mater. Chem. A*, 2015, **3**, 19353–19359.

18 F. Behrouznejad, S. Shahbazi, N. Taghavinia, H.-P. Wu and E. Wei-Guang Diau, *J. Mater. Chem. A*, 2016, **4**, 13488–13498.

19 Y. Kato, L. K. Ono, M. V. Lee, S. Wang, S. R. Raga and Y. Qi, *Adv. Mater. Interfaces*, 2015, **2**, 1500195.

20 C. Besleaga, L. E. Abramiuc, V. Stancu, A. G. Tomulescu, M. Sima, L. Trinca, N. Plugaru, L. Pintilie, G. A. Nemnes, M. Iliescu, H. G. Svavarsson, A. Manolescu and I. Pintilie, *The journal of physical chemistry letters*, 2016, **7**, 5168–5175.

21 Y. Han, S. Meyer, Y. Dkhissi, K. Weber, J. M. Pringle, U. Bach, L. Spiccia and Y.-B. Cheng, *J. Mater. Chem. A*, 2015, **3**, 8139–8147.

22 J. Li, Q. Dong, N. Li and L. Wang, *Adv. Energy Mater.*, 2017, **7**, 1602922.

23 J. You, L. Meng, T.-B. Song, T.-F. Guo, Y. M. Yang, W.-H. Chang, Z. Hong, H. Chen, H. Zhou, Q. Chen, Y. Liu, N. de Marco and Y. Yang, *Nature nanotechnology*, 2016, **11**, 75–81.

24 Z. Jiang, X. Chen, X. Lin, X. Jia, J. Wang, L. Pan, S. Huang, F. Zhu and Z. Sun, *Solar Energy Materials and Solar Cells*, 2016, **146**, 35–43.

25 K. Domanski, J.-P. Correa-Baena, N. Mine, M. K. Nazeeruddin, A. Abate, M. Saliba, W. Tress, A. Hagfeldt and M. Grätzel, *ACS nano*, 2016, **10**, 6306–6314.

26 S. Cacovich, L. Ciná, F. Matteocci, G. Divitini, P. A. Midgley, A. Di Carlo and C. Ducati, *Nanoscale*, 2017, **9**, 4700–4706.

27 R. A. Kerner, P. Schulz, J. A. Christians, S. P. Dunfield, B. Dou, L. Zhao, G. Teeter, J. J. Berry and B. P. Rand, *APL Materials*, 2019, **7**, 41103.

28 D. R. Lide, *CRC handbook of chemistry and physics, 1998-1999*, CRC Press, Boca Raton, 79th edn., 1998.

29 W. Sun, S. Ye, H. Rao, Y. Li, Z. Liu, L. Xiao, Z. Chen, Z. Bian and C. Huang, *Nanoscale*, 2016, **8**, 15954–15960.

30 B. P. Rand, J. Li, J. Xue, R. J. Holmes, M. E. Thompson and S. R. Forrest, *Adv. Mater.*, 2005, **17**, 2714–2718.

31 J. Xue, B. P. Rand, S. Uchida and S. R. Forrest, *Journal of Applied Physics*, 2005, **98**, 124903.

32 Q. Jiang, X. Sheng, B. Shi, X. Feng and T. Xu, *J. Phys. Chem. C*, 2014, **118**, 25878–25883.

33 J. A. Christians, R. C. M. Fung and P. V. Kamat, *Journal of the American Chemical Society*, 2014, **136**, 758–764.

34 I. Zarazua, G. Han, P. P. Boix, S. Mhaisalkar, F. Fabregat-Santiago, I. Mora-Seró, J. Bisquert and G. Garcia-Belmonte, *The journal of physical chemistry letters*, 2016, **7**, 5105–5113.

35 F. Fabregat-Santiago, G. Garcia-Belmonte, I. Mora-Seró and J. Bisquert, *Physical chemistry chemical physics : PCCP*, 2011, **13**, 9083–9118.

36 A. Todinova, L. Contreras-Bernal, M. Salado, S. Ahmad, N. Morillo, J. Idígoras and J. A. Anta, *ChemElectroChem*, 2017, **4**, 2891–2901.

37 X. Xu, Z. Liu, Z. Zuo, M. Zhang, Z. Zhao, Y. Shen, H. Zhou, Q. Chen, Y. Yang and M. Wang, *Nano letters*, 2015, **15**, 2402–2408.

38 H. T. G. Hentzell, C. R. M. Grovenor and D. A. Smith, *Journal of Vacuum Science & Technology A: Vacuum, Surfaces, and Films*, 1984, **2**, 218–219.

39 C.R.M. Grovenor, H.T.G. Hentzell and D. A. Smith, *Acta Metallurgica*, 1984, **32**, 773–781.

40 B. A. Movchan and A. V. Demchishin, *Phys. Met. Metallogr. USSR*, 1969, **28**, 83.

41 J. A. Thornton, *Journal of Vacuum Science and Technology*, 1975, **12**, 830–835.

42 J. A. Thornton, *Journal of Vacuum Science and Technology*, 1974, **11**, 666–670.

43 M. Ohring, *The materials science of thin films*, Academic Press, Boston, 1992.

44 R. W. Balluffi and J. M. Bkakely, *Thin Solid Films*, 1975, **25**, 363–392.

45 N. Ahn, D.-Y. Son, I.-H. Jang, S. M. Kang, M. Choi and N.-G. Park, *Journal of the American Chemical Society*, 2015, **137**, 8696–8699.

46 J.-W. Lee, H.-S. Kim and N.-G. Park, *Accounts of chemical research*, 2016, **49**, 311–319.

# Supplementary information

1. Cross-section SEM microphotographs of aged PSCs with different contact materials

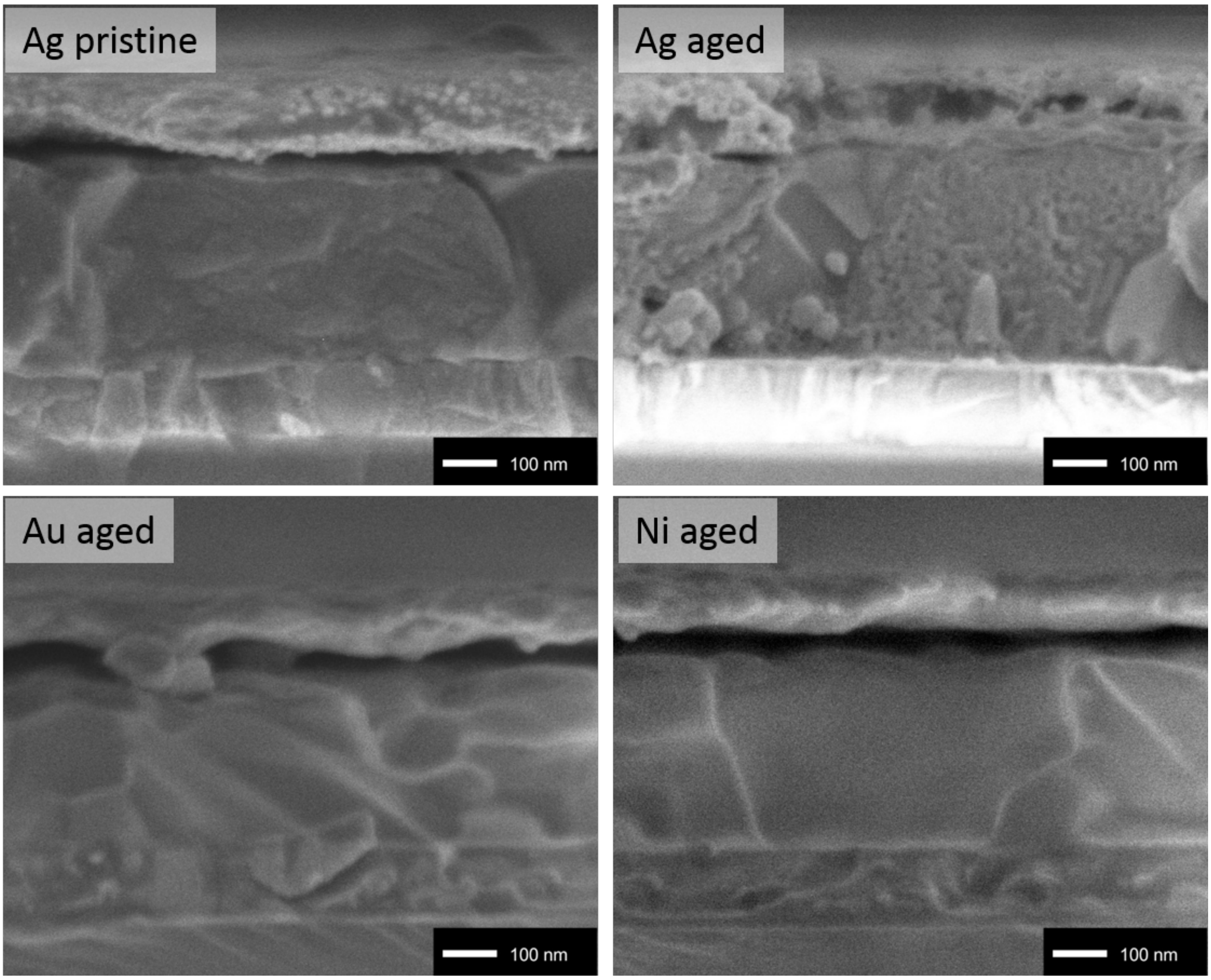

**Figure 1. Cross section SEM microphotographs of pristine PSC with Ag and aged PSCs with Ag, Au and Ni contacts.**

## 2. Adhesion tape test

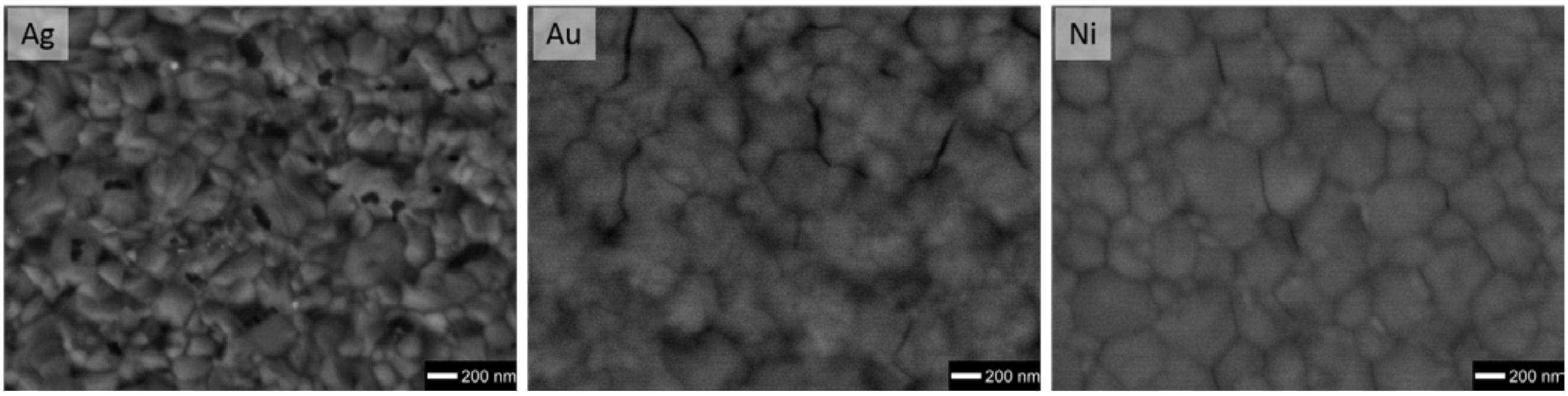

**Figure 2. Adhesion test of Ag, Au and Ni contacts on the top of MAPI/C60/BCP after 2-week storage in air.**

## 3. IS measurements

Following equations were used for fitting the impedance measurements:

$$Z' = R_{ser} + \frac{R_{per/ETL}}{1 + R_{per/ETL}^2 \omega^2 C_{per/ETL}^2}$$

$$Z'' = -j \frac{R_{per/ETL}^2 \omega C_{per/ETL}}{1 + R_{per/ETL}^2 \omega^2 C_{per/ETL}^2}$$

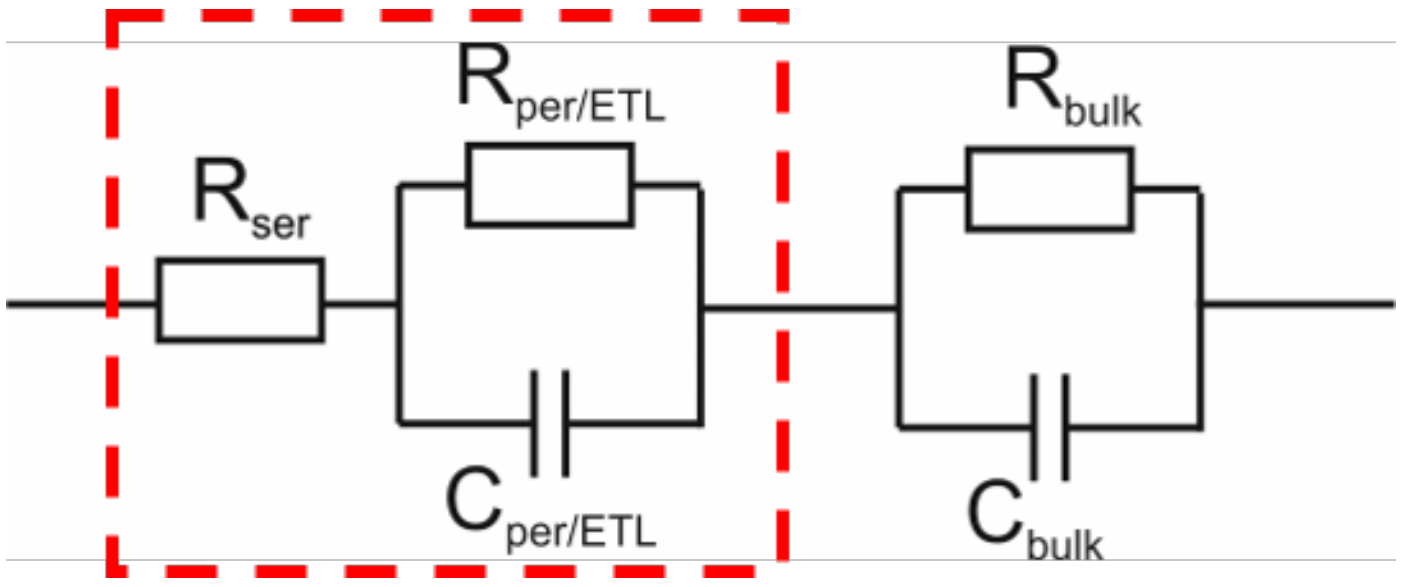

**Figure 3. Equivalent electrical circuit for IS measurements**

**Table 1. Fitting parameter for IS measurements**

| | Au | Ni | Ag |
|---|---|---|---|
| pristine | $R_{ser}$ = 15.1 Ω<br>$C_{per/ETL}$ = 2.1E-8 F<br>$R_{per/ETL}$ = 40.7 Ω | $R_{ser}$ = 11.7 Ω<br>$C_{per/ETL}$ = 1.1E-8 F<br>$R_{per/ETL}$ = 35.1 Ω | $R_{ser}$ = 15.9 Ω<br>$C_{per/ETL}$ = 1.1E-8 F<br>$R_{per/ETL}$ = 83.9 Ω |
| 2-week air storage | $R_{ser}$ = 29.2 Ω<br>$C_{per/ETL}$ = 2.4E-8 F<br>$R_{per/ETL}$ = 248.9 Ω | $R_{ser}$ = 19.6 Ω<br>$C_{per/ETL}$ = 1.1E-8 F<br>$R_{per/ETL}$ = 72.8 Ω | $R_{ser}$ = 156.5 Ω<br>$C_{per/ETL}$ = 5.8E-9 F<br>$R_{per/ETL}$ = 1745,7 Ω |

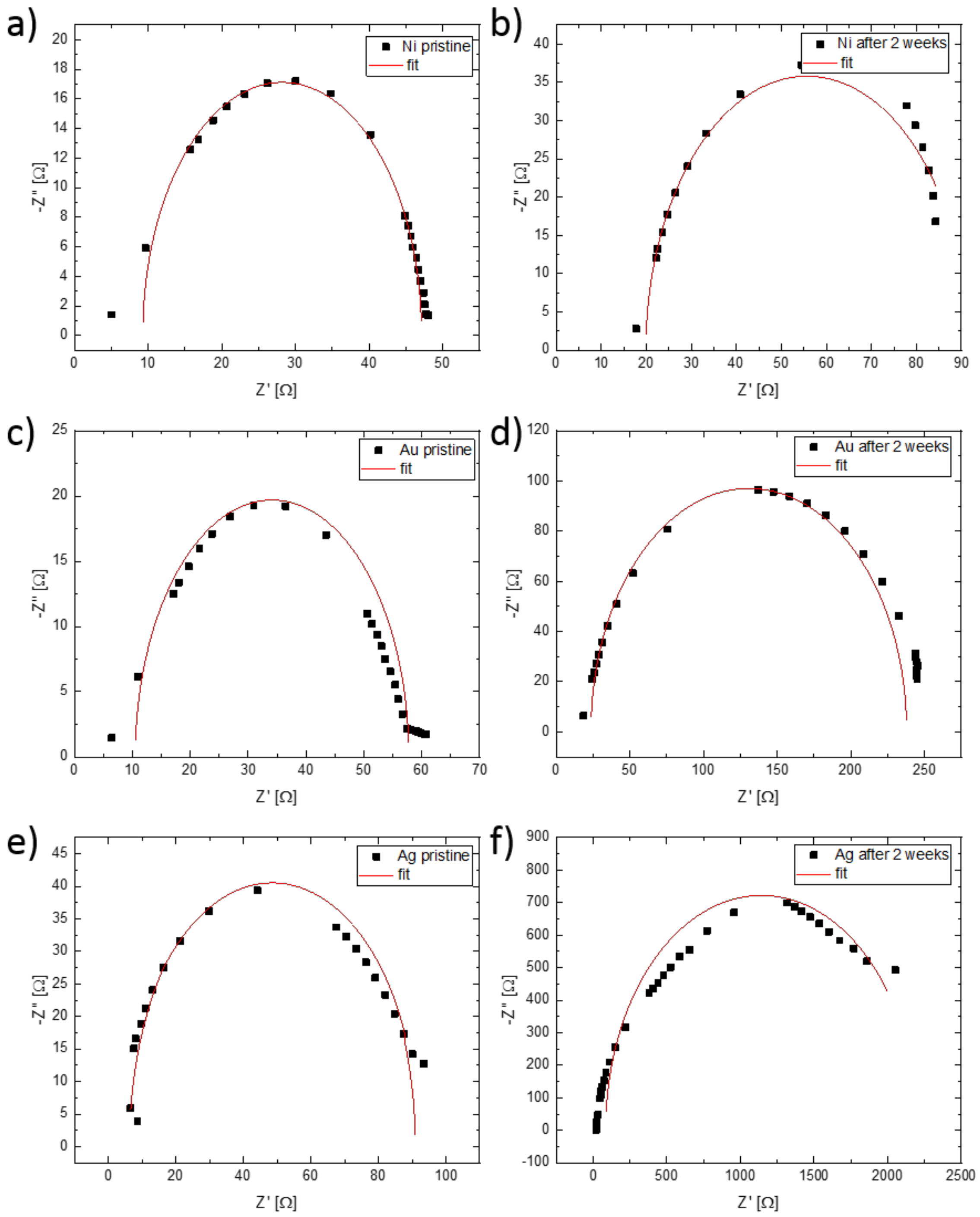

**Figure 4. IS measurements at pristine (a, c, e) and 2-week old (b, d, f) PSCs with Ag (a,b), Au (c, d) and Ni (e,f) contacts.**

4. EDX mapping of cross-section of pristine and two-month-old devices.

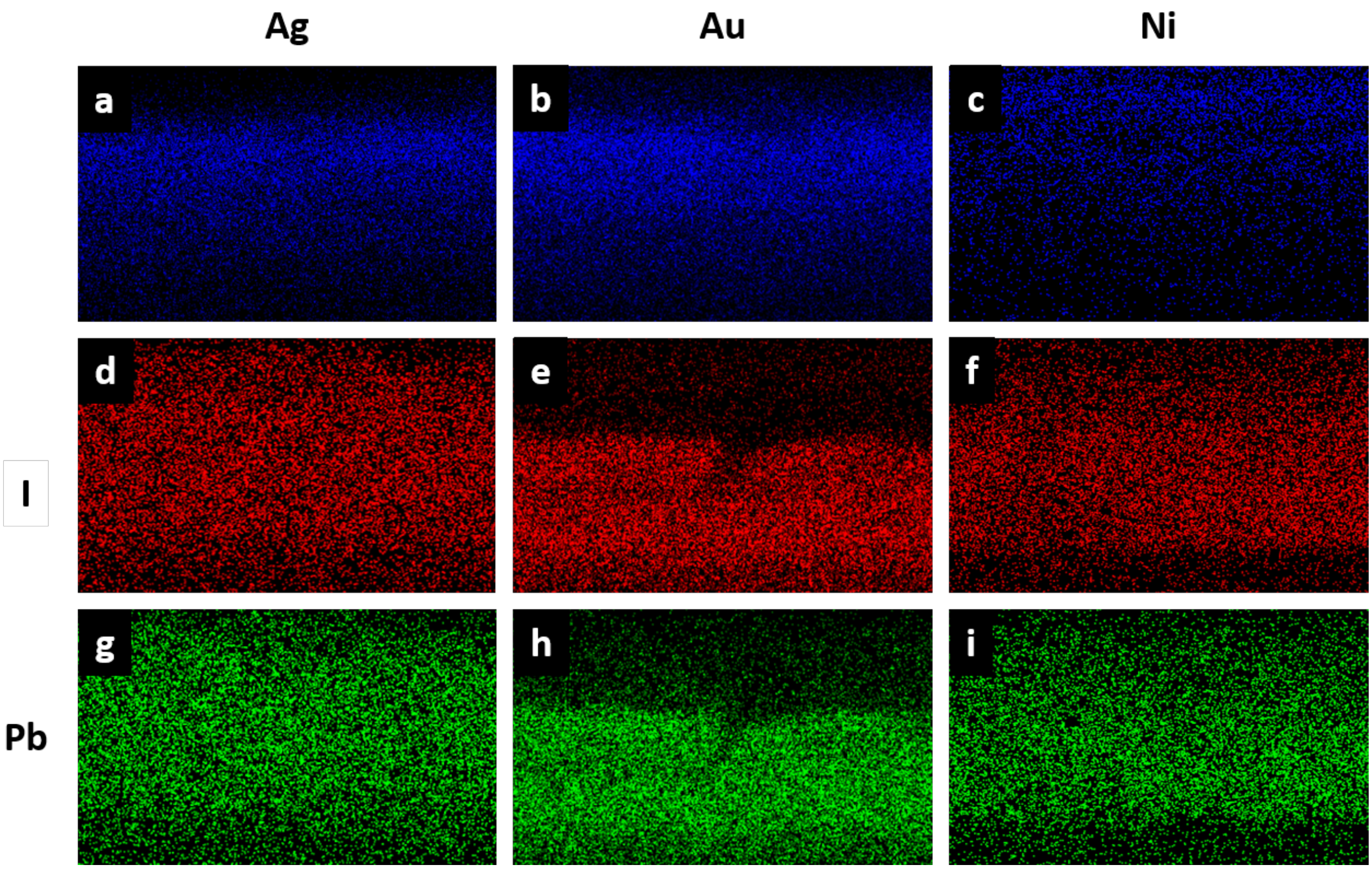

**Figure 6. Cross section EDX mapping of pristine PSCs with Ag, Au and Ni contacts. Metals, iodide and lead are presented with blue, red and green colors respectively.**

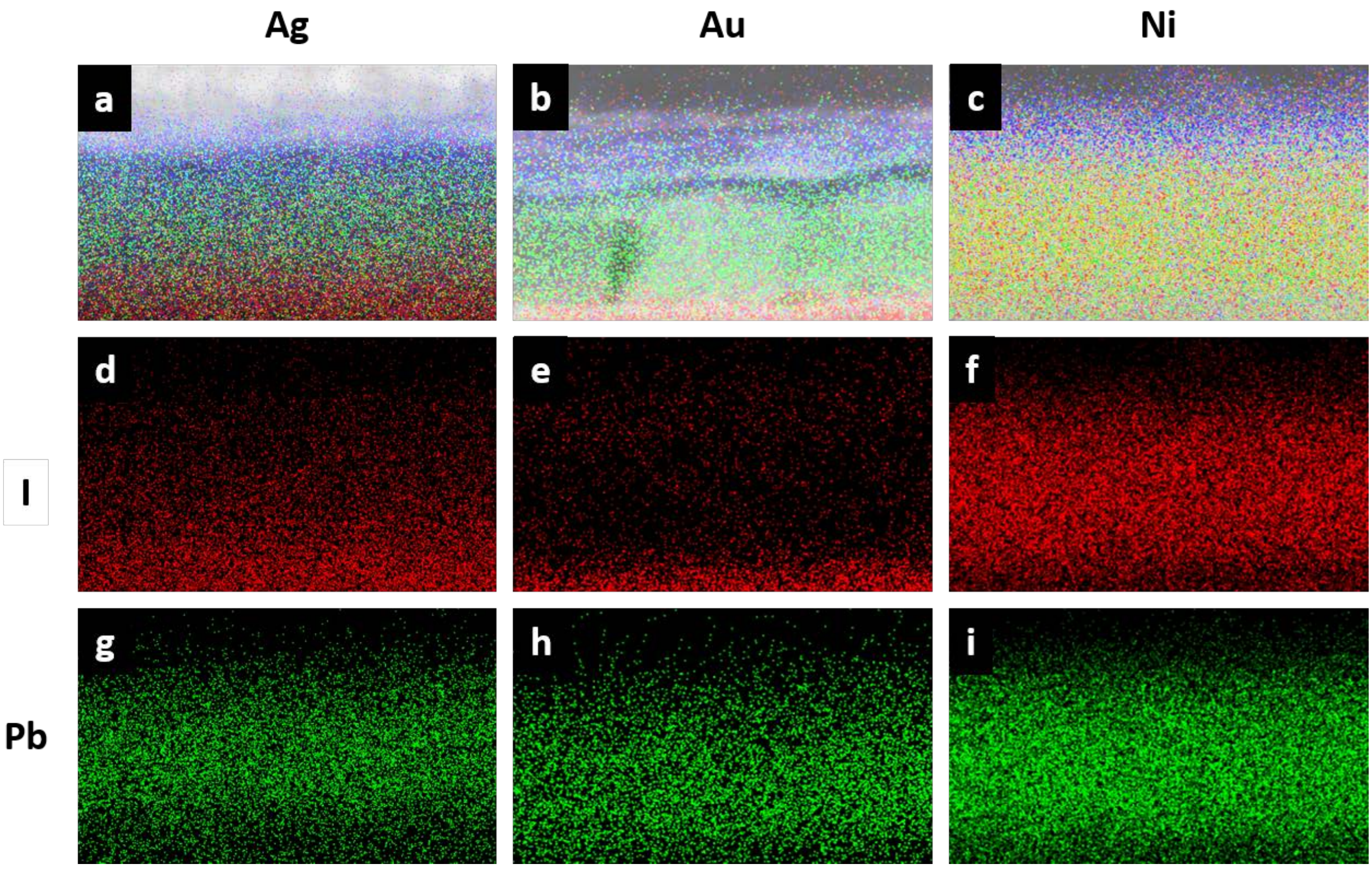

**Figure 5. Cross section EDX mapping of 2-month old PSCs with Ag, Au and Ni contacts. Metals, iodide and lead are presented with blue, red and green colors respectively.**